\documentclass[twocolumn,showpacs,preprintnumbers,amsmath,amssymb,prb,superscriptaddress]{revtex4}

\usepackage{graphicx}

\newcommand{\expct}[2]{\left\langle #1 \right\rangle_{#2}}
\newcommand{\expcts}[2]{\langle #1 \rangle_{#2}}
\newcommand{\C}{\mathcal{C}}
\newcommand{\TC}{T_{\C}\, }

\newcommand{\shat}[1]{\smash{\hat{#1}}}

\newcommand{\real}{\text{Re}}
\newcommand{\imag}{\text{Im}}

\begin{document}

\title{Transient dynamics of a molecular quantum dot with a vibrational degree of freedom}
\author{R.-P.~Riwar}
\affiliation{Department of Physics, University of Basel,
Klingelbergstrasse 82, CH-4056 Basel, Switzerland}
\affiliation{Institut f\"{u}r Theoretische Physik A, RWTH Aachen, D-52056 Aachen, Germany}
\author{T.~L.~Schmidt}
\affiliation{Department of Physics, University of Basel,
Klingelbergstrasse 82, CH-4056 Basel, Switzerland}
\affiliation{Department of Physics, Yale University, 217 Prospect Street, New Haven, CT 06520, USA}
\date{\today}

\begin{abstract}
We investigate the transient effects occurring in a molecular quantum dot described by an Anderson-Holstein Hamiltonian which is instantly coupled to two fermionic leads biased by a finite voltage. In the limit of weak electron-phonon interaction, we use perturbation theory to determine the time-dependence of the dot population and the average current. The limit of strong coupling is accessed by means of a self-consistent time-dependent mean-field approximation. These complementary approaches allow us to investigate the dynamics of the inelastic effects occurring when the applied bias voltage exceeds the phonon frequency and the emergence of bistability.
\end{abstract}

\pacs{73.23.-b, 72.10.Di}

\maketitle

\section{Introduction}

In recent years, interest in the interplay of electronic and vibrational degrees of freedom of individual electrically contacted molecules has given rise to the prospering field of molecular electronics.\cite{galperin07,cuniberti05,nitzan03,aviram74} Single-molecule transistors\cite{park00} and memory cells\cite{reed01} have been proposed and the investigation of phonon spectra of single molecules by an electric measurement has become possible. The two most prominent experimental setups for contacting individual molecules are STM tips\cite{stipe98,zhitenev02,qiu04} and mechanically controllable break junctions\cite{smit02,djukic05}. These were used successfully to measure the influence of vibrational degrees of freedom on current and noise properties of systems as distinct as hydrogen,\cite{smit02,djukic05,djukic06} water\cite{tal08} and benzene\cite{kiguchi08} molecules, fullerenes,\cite{yu04_2,pasupathy05,parks07} as well as carbon nanotubes.\cite{leroy04,tans97,sapmaz06}

It was found that the vibrations of the molecule can indeed have discernible effects on conductance properties. One of the most striking of these is an abrupt increase or decrease (depending on the transmission) in the differential conductance once the applied voltage exceeds a vibration frequency and thus allows for the excitation of a phonon.\cite{delavega06,frederiksen04,paulsson08} Recently, it was shown that similar features at this threshold can appear in the shot noise properties.\cite{galperin06,schmidt09,avriller09,haupt09} This is the reason why measurements of transport properties are becoming an invaluable tool for the determination of phonon spectra of single molecules. Such effects have already been observed in a number of experiments and do not require a strong electron-phonon coupling.\cite{egger08,paulsson05}

In the opposite limit of strong electron-phonon coupling, even more spectacular effects were predicted. These include the suppression of sequential tunneling due to the Franck-Condon effect\cite{koch05} or the emergence of bistability. The latter has been argued to lead to switching between two stable configurations of the molecule and to hysteresis effects.\cite{gogolin02,ryndyk08,damico08,galperin08,loertscher06,alexandrov03}

Even from the theoretical point of view, the knowledge of these systems is still far from complete. Generally, the electron-phonon coupling on the molecule leads to the formation of a local polaron.\cite{mahan90} While the Hamiltonian of the isolated molecule can be diagonalized in terms of these polarons, no exact solution is known in the presence of tunneling to the contacts, such that various approximate numerical and analytical schemes have been developed. In the case of small electron-phonon coupling, perturbation theory has proved successful.\cite{mitra04,egger08,schmidt09,haupt09,avriller09} In the case of small tunneling, results have been obtained using rate equations,\cite{flindt05,chtchelkatchev04} perturbation theory\cite{leijnse08} or a truncation scheme within a Green's function (GF) formalism, leading to a mean-field approximation.\cite{ryndyk08,damico08}

In order to achieve a broader understanding of this system, and in particular of the phonon excitation process and the emergence of bistability, we shall investigate the time evolution of a molecular quantum dot with a single vibrational mode as a reaction to a sudden switching-on of the coupling to the leads. Time-dependent transport properties of mesoscopic systems have been examined in various contexts\cite{vanevic08,moskalets08} and, in particular, such switching effects have been investigated previously for the case of a noninteracting quantum dot,\cite{jauho94,maciejko06} and for a weakly Coulomb interacting quantum dot.\cite{schmidt08} Strongly interacting systems in the Kondo regime have been investigated in Refs.~[\onlinecite{nordlander99,plihal05,goker07,komnik09}]. Generally, it was shown that the time evolution is governed by the tunneling amplitude but non-adiabatic effects depending on various other system parameters were also predicted. Experimental measurements of such transients have been achieved using quantum point contact setups.\cite{vink07}

The structure of this paper is as follows: In Section II, we shall introduce the model used to describe the molecular quantum dot, the coupling to the phonon and the implementation of the switching. In Section III, the results for the case of weak electron-phonon interaction will be derived using perturbation theory. In Section IV, we go to the opposite limit of strong electron-phonon interaction and present results obtained by means of a self-consistent time-dependent mean-field approximation. The two results will be compared in the appropriate limits and our conclusions will be detailed in Section V.

\section{System}

In general, the typical systems investigated in the field of molecular electronics can be rather complicated: the Coulomb interaction between electrons on the dot may be quite strong, complicated molecules may support a whole spectrum of phonon excitations and the transport of electrons to the leads may occur via multiple channels of different transparencies.

In order to avoid these complications, we focus on the archetype model which is able to capture the most prominent effects in the physics of molecular junctions. It is described by the Anderson-Holstein Hamiltonian
\begin{equation}\label{H}
H=\sum_{\alpha=L,R}H_\alpha+H_{\text{dot}}+H_{\text{ph}}+H_{\text{el-ph}}+H_T
\end{equation}
and it has been widely used in the literature.
\cite{galperin07, egger08, damico08} The left and right lead Hamiltonians $H_{L,R}$ describe noninteracting electron gases with the chemical potentials $\mu_\alpha$, which can be tuned by the applied bias voltage $V = \mu_L - \mu_R$, (we use units where $e = \hbar = 1$)
\begin{equation}
H_\alpha=\sum_k(\epsilon_k-\mu_\alpha)c^{\dagger}_{k,\alpha}c_{k,\alpha}\ .
\end{equation}
Without loss of generality, we shall assume symmetric bias voltage $\mu_{L,R} = \pm V/2$ in the following. The second term in the Hamiltonian (\ref{H}) represents a single level quantum dot with energy $\Delta$,
\begin{equation}
H_{\text{dot}}=\Delta d^\dagger d\ ,
\end{equation}
which can correspond to either the LUMO or HOMO energy level of the molecule depending on the ground state population. For simplicity, we have discarded the electron spin. The interaction of the electron on the dot with a Holstein phonon of frequency $\Omega$, described by $H_{ph}=\Omega a^\dagger a$, is included up to linear order in the oscillator displacement $q \sim a + a^\dag$ via the electron-phonon coupling parameter $\lambda$
\begin{equation}
H_{\text{el-ph}}=\lambda (a+a^\dagger) d^\dagger d\ .
\end{equation}
The connection between the dot and the leads is described by a local tunneling Hamiltonian with amplitudes $\gamma_{L,R}$. In order to observe switching effects, we allow for time-dependent amplitudes and thus use
\begin{equation}\label{HT}
H_T(t) = \sum_\alpha \gamma_\alpha(t) \Big[\psi^\dagger_\alpha(x=0) d+\text{h.c.} \Big]\ , 
\end{equation}
where the electron field operator $\psi_\alpha(x)$ is the Fourier transform of the electron annihilation operator $c_{k,\alpha}$. The assumption of local tunneling at $x=0$ allows us to discard the spatial variable $x$, ie.~we shall use $\psi_{L,R} \equiv \psi_{L,R}(x=0)$. As to the time dependence, we shall assume that the tunneling is instantly switched on at $t=0$ which means $\gamma_\alpha(t) = \gamma_\alpha \theta(t)$, where $\theta(t)$ denotes the Heaviside step-function. In order to simplify the calculation, we shall assume a spatially symmetric system, i.e.~$\gamma_L = \gamma_R = \gamma$ and restrict ourselves to zero temperature in the following.

We are mainly interested in the time evolution of the dot population $\hat{n}(t) = d^\dag(t) d(t)$ and the time-dependent current through the left/right contact. Using the Heisenberg equation of motion, the current operator $\shat{I}_\alpha$ ($\alpha = L,R$) can be calculated as the time derivative of the total charge $\shat{Q}_\alpha$ in the respective lead,
\begin{equation}
\hat{I}_{L,R} = \mp \frac{d Q_{L,R}}{dt} = \pm i \gamma \left( \psi^\dag_{L,R} d - d^\dag \psi_{L,R} \right)\ .
\end{equation} 
The signs are chosen such that current flowing from left to right will be positive. The expectation values $I_\alpha(t) = \expcts{\shat{I}_\alpha(t)}{}$ and $n(t) = \expcts{\shat{n}(t)}{}$ can be expressed in terms of the dot Keldysh GF,
\begin{equation}
D(t,t') = -i \expct{\TC d(t) d^\dag(t')}{}\ ,
\end{equation} 
where $\TC$ denotes the time-ordering operator on the Keldysh contour $\C$ and the time variables $t$ and $t'$ can be situated on the forward or backward part of the Keldysh contour $\C_\pm$, giving rise to the GF matrix,
\begin{equation}
{\bf D} = \left[ \begin{array}{cc}
D^{--} & D^{-+} \\
D^{+-} & D^{++}\end{array} \right]\ .
\end{equation}
For our purposes it turns out to be more convenient to introduce the retarded and advanced GFs, which are defined by
\begin{align}
D^R &= D^{--} - D^{-+} \notag \\
D^A &= D^{+-} - D^{++}\ .
\end{align}
Now, the time-dependent dot occupation can be expressed as
\begin{equation}\label{n}
n(t) = -i D^{-+}(t,t)\ ,
\end{equation} 
while the current can be written as a sum of two contributions, $I_\alpha = I_\alpha' + I_\alpha''$, which are defined as\cite{schmidt08}
\begin{align}
I'_{L,R}(t) &= \mp \gamma^2\ \real \int_{0}^{\infty} dt_{1}\ g_{L,R}^R(t,t_{1})\ D^{K}(t_{1},t)\ \label{I1} , \\
I''_{L,R}(t) &= \pm \gamma^2\ \real \int_{0}^{\infty}dt_{1}\
D^R(t,t_{1})\ g_{L,R}^K(t_{1},t)\ \label{I2} .
\end{align}
Here, $g_\alpha(t,t') = -i \expcts{\TC \psi_\alpha(t) \psi_\alpha^\dag(t')}{0}$ denotes the uncoupled lead GF, where the expectation value is taken with respect to the ground state of the uncoupled Hamiltonian $H_L + H_R$. It is given by
\begin{eqnarray}
 {\bf g}_\alpha(\omega) = i 2 \pi \rho(\omega) \left[
 \begin{array}{cc}
 f_\alpha - 1/2 & f_\alpha \\
-(1-f_\alpha) & f_\alpha - 1/2
\end{array} \right] \, ,
\end{eqnarray}
which depends on the density of states $\rho(\omega)$ in the leads and the Fermi functions $f_\alpha(\omega) = n_F(\omega - \mu_\alpha)$. The superscript ``$K$'' in Eqs.~(\ref{I1}) and (\ref{I2}) denotes Keldysh GFs which are defined by $D^K = D^{-+} + D^{+-} = 2 D^{-+} + D^R - D^A$ and analogously for $g^K$. The currents through the two contacts are not independent since the symmetric bias entails $I_L(V) = - I_R(-V)$.

In the noninteracting case ($\lambda = 0$), the GF $D_0(t,t')$ can be calculated analytically even for the time-dependent tunneling in Eq.~(\ref{HT}). The calculation is greatly simplified by assuming a constant density of states $\rho_0$ in both leads. Introducing the contact tunnelling rate $\Gamma = 2 \pi \rho_0 \gamma^2$, one finds\cite{schmidt08,langreth91} for $t,t' > 0$,
\begin{align}\label{D0RA}
 D^{(0)R}(t-t') &= - i \theta(t-t') \theta(t') \, e^{ - i \Delta (t-t')} \, e^{ -
 \Gamma (t-t')} \notag \\
 D^{(0)A}(t-t') &= i \theta(t'-t) \theta(t) \, e^{ - i \Delta (t-t')} \, e^{ 
 \Gamma (t-t')}
\end{align}
and
\begin{align}\label{D0mp}
D^{(0)-+}(t,t')&= \frac{i \Gamma}{2\pi}\theta(t) \theta(t') \sum_{\alpha = L,R} \int_{-\infty}^{\infty} d\omega\frac{f_{\alpha}(\omega+\Delta)}{\Gamma^2+\omega^2} \notag \\
&\times \big(e^{-i\omega t}-e^{-\Gamma t}\big)\big(e^{i\omega t'}-e^{-\Gamma t'}\big)\ .
\end{align}
This leads to the following evolution of the dot occupation number in the case of an initially empty dot $n(0) = 0$,
\begin{eqnarray} \label{n0t}
n^{(0)}(t)
&=& 
 \frac{\Gamma}{\pi} \theta(t) e^{-\Gamma t} 
 \sum_\alpha \int_{-\infty}^{\infty} d\omega \frac{f_{\alpha}(\omega+\Delta)}{\Gamma^2+\omega^2} \\ \nonumber 
&\times& 
 \left[ \cosh(\Gamma t)- \cos(\omega t)\right]\ .
\end{eqnarray}
The time-dependent current can be split into a displacement current $I_{\text{disp}}(t) = I_L(t) - I_R(t)$, which reflects the change in dot population, $I_{\text{disp}}(t) = -\dot{n}(t)$, and the total current $I(t) = [I_L(t) + I_R(t)]/2$ which measures the charge transported through the system. Both current terms as well as the dot population $n^{(0)}(t)$ where calculated in [\onlinecite{schmidt08}] for the non-interacting case. Therefore, in the following sections, we will focus on how these expressions change due to the presence of electron-phonon interaction.

\section{Perturbation theory in electron-phonon coupling}

In the limit of weak electron-phonon interaction, we treat the system perturbatively in $\lambda$. The two non-vanishing corrections up to order $\lambda^2$ are commonly referred to as the ``tadpole'' (subscript 1) and the ``rainbow'' term (subscript 2). The former represents the interaction of a phonon with the electron density, whereas the latter describes virtual phonon mode exitations. The corresponding dot GF contributions are (superscripts $(2)$ denote expressions to second order in $\lambda$)
\begin{align}\label{D2tadpole}
D^{(2)}_1(s,s') &= 2\lambda^2 \int_\C ds_{1} \int_\C ds_2\ D^{(0)}(s,s_{1}) \\ \notag
&\times D^{(0)}(s_{1},s')
F(s_{1}-s_{2}) n^{(0)}(s_{2})
\end{align}
and
\begin{align}\label{D2rainbow}
D^{(2)}_{2}(s,s') &=
2i\lambda^2 \int_\C ds_{1} \int_\C ds_{2}\ D^{(0)}(s,s_{1}) \\ \notag
&\times D^{(0)}(s_{1},s_{2}) F(s_{1}-s_{2}) D^{(0)}(s_{2},s')
\end{align}
The time integrations run along the Keldysh contour $\C$ (where all time variables are on either the forward or backward part of the time loop) and $F(t-t')$ denotes the unperturbed Keldysh GF of the phonon. In the case of zero temperature, it is given by
\begin{equation}
{\bf F}(t-t') =-\frac{i}{2}\left[\begin{array}{cc}
e^{-i\Omega|t-t'|} & e^{i\Omega(t-t')}\\
e^{-i\Omega(t-t')} & e^{i\Omega|t-t'|}\end{array}\right].
\end{equation}
Both diagrams translate into time-dependent contributions to the dot occupation number which are given by $\smash{n^{(2)}_{1,2}(t) = -i D^{(2)-+}_{1,2}(t,t)}$. Up to the second order in the electron-phonon interaction, the total dot population is given by $\smash{n(t) = n^{(0)} + n^{(2)}_1 + n^{(2)}_2}$. One can easily derive the following expressions for the second order contributions. For the tadpole term, one finds
\begin{align}\label{n2tadpole}
 n^{(2)}_1(t) &= 4 \lambda^2\ \imag \Bigg[ \int_0^t dt_1\  D^{(0)R}(t-t_1) D^{(0)-+}(t_1,t) \notag \\
&\times \int_0^{t_1} dt_2\ F^{R}(t_1-t_2) n^{(0)}(t_2) \Bigg]
\end{align}
where we used the retarded phonon GF $\smash{F^R(t-t') = -i \theta(t-t') \sin [ \Omega (t-t') ]}$. Moreover, the contribution from the rainbow term becomes
\begin{widetext}
\begin{align}\label{n2rainbow}
 n^{(2)}_2(t) &=
4 \lambda^2\ \real \int_0^{t} dt_1 \int_0^{t_1} dt_2 D^{(0)R}(t-t_1)
 \left[ D^{(0)-+}(t_1,t_2) F^R(t_1-t_2) + D^{(0)R}(t_1-t_2) F^{--}(t_1-t_2) \right] D^{(0)-+}(t_2,t) \notag \\
&+ 2 \lambda^2 \int_0^t dt_1 \int_0^t dt_2 D^{(0)R}(t,t_1) D^{(0)-+}(t_1,t_2) F^{-+}(t_1-t_2) D^{(0)A}(t_2,t)\ .
\end{align}
\end{widetext}
In order to calculate the time-dependent current, we make use of the general Eqs.~(\ref{I1}) and (\ref{I2}) and write the second order current as $I^{(2)}_\alpha = I'^{(2)}_\alpha + I''^{(2)}_\alpha$. Using the unperturbed retarded GF $g^R_{L,R}(t) = -i \pi \rho_0 \delta(t)$, one finds for the first term $\smash{I'_{L,R}(t) = \pm \theta(t) \tfrac{\Gamma}{2} [ 1 - 2 n(t) ]}$, so the corresponding second order contribution reads
\begin{align}
 I'^{(2)}_{L,R}(t) = \mp \Gamma [ n^{(2)}_1(t) + n^{(2)}_2(t)]
\end{align}
Due to its symmetry, this term does not create any net current $I(t)$ but only contributes to the displacement current $I_{\text{disp}}(t)$. The second contribution from Eq.~(\ref{I2}) is given by
\begin{align}
 I''^{(2)}_{L,R}(t) &= \mp \Gamma\ \imag \Bigg\{ \int_{0}^{t} dt_1 
\left[D^{(2)R}_1(t,t_1) + D^{(2)R}_2(t,t_1) \right] \notag \\
&\times \int \frac{d\omega}{2\pi} e^{i \omega (t_1 - t)} [2 f_{L,R}(\omega) - 1] \Bigg\}
\end{align}
For the evaluation of these observables, we shall focus on the particle-hole symmetric case, $\Delta = 0$. Deviations from this point were investigated in detail in Ref.~[\onlinecite{schmidt08}] and it was shown that they can have a clear influence on the time-dependent current and dot population as they cause oscillatory behavior on a time scale $\Delta^{-1}$. As these oscillations might obscure features induced by the coupling to the phonon, we choose to set $\Delta = 0$. The remaining integrals can easily be solved by a simple numerical integration which eventually leads to the time-dependent current and dot occupation.

\begin{figure}[t]
  \centering
  \includegraphics[width = 0.48 \textwidth]{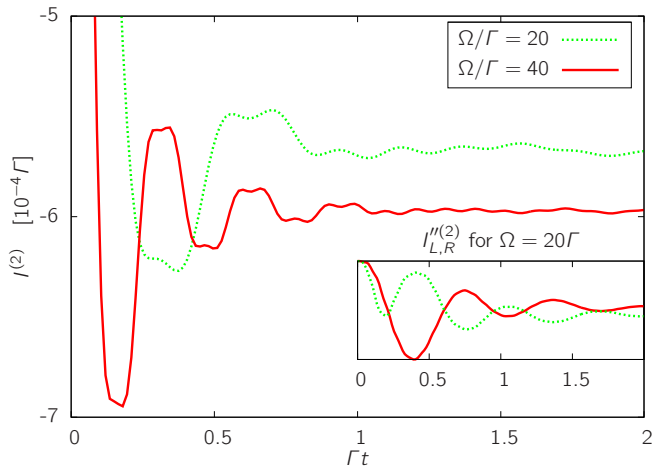}
  \caption{(Color online) Time-dependent current $I^{(2)}(t)$ for fast phonon mode $\Omega \gg \Gamma$. A retardation of order $\Omega^{-1}$ between currents across left and right contact leads to the emergence of plateaus in the time trace of the total current. Parameters are $V = \Omega, \lambda = \Omega/40, \Delta = 0$.}
  \label{fig:plateau}
\end{figure}

For a fast phonon mode, in the limit $\Gamma \ll \Omega$, the total time-dependent current $\smash{I^{(2)} = [I^{(2)}_L + I^{(2)}_R]/2}$ exhibits plateaus (as depicted in Fig.~\ref{fig:plateau}) that originate from a retardation between left and right current, $\smash{I^{(2)}_L(t)}$ and $\smash{I^{(2)}_R(t)}$. Since both the width of the plateaus and, naturally, their period are proportional to $\Omega^{-1}$, we attribute this to an effect comparable to electron shuttling,\cite{novotny04,isacsson04,flindt05} in this particular case due to the electron-phonon interaction.

\begin{figure}[t]
  \centering
  \includegraphics[width = 0.48 \textwidth]{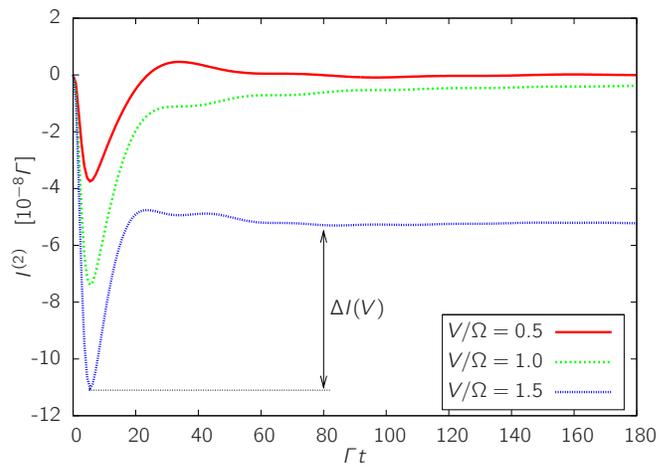}
  \caption{(Color online) Time-dependent current $I^{(2)}(t)$ for slow phonon mode $\Omega \ll \Gamma$. Shortly after switching on the tunneling, the current overshoots to values beyond the steady-state value. The overshooting amplitude is characterized by $\Delta I(V)$. Parameters are $\Omega = 0.1 \Gamma$, $\lambda = \Omega/40$, $\Delta = 0$.}
  \label{fig:Itot}
\end{figure}

\begin{figure}[t]
  \centering
  \includegraphics[width = 0.48 \textwidth]{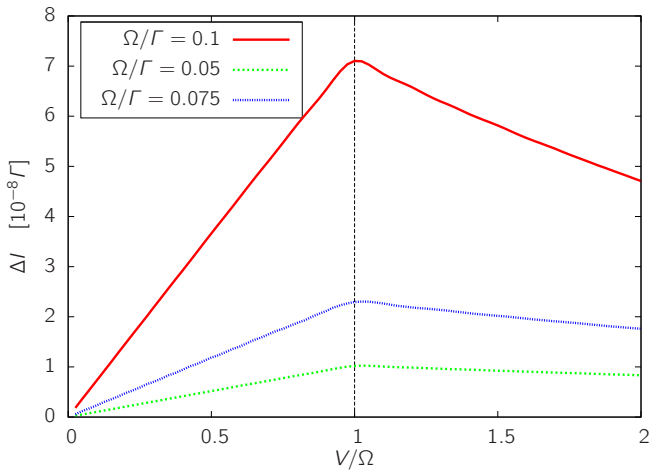}
  \caption{(Color online) Overshooting strength $\Delta I$ as a function of bias voltage $V$. The overshooting displays a maximum at the threshold for inelastic processes.}
  \label{fig:Iovershoot}
\end{figure}

When going to the opposite regime of low phonon frequency, $\Gamma\gg\Omega$, there is a significant overshoot of the total current correction for short times (see Fig.~\ref{fig:Itot}). Scanning through different voltages, one sees that the negative peak of the perturbative current grows linearly with $V$. We can characterise the relative strength of the overshoot by the difference between the (negative) current peak and its stationary value as a function of voltage
\begin{align}
\Delta I (V) =  \lim_{t\rightarrow\infty} [I^{(2)}(t)] - \min_t [I^{(2)}(t)],
\end{align}
which results in Fig.~\ref{fig:Iovershoot}. The relative overshoot still increases linearly for $V < \Omega$ but begins to decrease thereafter. We know from the stationary-state calculation\cite{egger08} that at $V = \Omega$, the phonon mode can be excited and inelastic processes set in which in our parameter range (large transmission) lead to a decrease in the stationary current. This suggests the conclusion that the time-dependent current can also be split into an elastic part (which is continuous at $V = \Omega$) and an inelastic one (which vanishes identically for $V < \Omega$) and that the overshooting is dominated by the elastic part.

We would like to point out that in the first case of the fast phonon regime, the emergent plateaus are damped on the timescale of $\Gamma^{-1}$. In the $\Gamma\gg\Omega$ limit on the other hand, the timescale for the dominant effect, the overshooting, is also of the order of $\Gamma^{-1}$. Therefore, the plateau feature, which is dominant in the former regime, is not visible anymore, because the time evolution of the phonons is too slow.

\section{Mean-field approximation}

The perturbation theory is valid only for low electron-phonon coupling $\lambda$. In order to describe features like bistability, the electron-phonon coupling strength needs to be much higher. A possible route to a solution for strong interaction is provided by a Hartree-like mean-field ansatz, which we would like to justify from the following physical point of view.

We know from the non-interacting case in Eq.~(\ref{n0t}) that there are basically two time scales governing the time-evolution for instant switching-on of the tunneling. It is the exponential saturation on the scale $\Gamma^{-1}$ as well as an oscillation with a frequency of order $\Delta$. If the oscillator period $\Omega^{-1}$ is much shorter than any of these time scales ($\Omega\gg\Delta,\Gamma$) then the change of the dot population will look very slow from the oscillator's point of view, and the oscillator rest position can adapt adiabatically to the momentary $n(t)$. Therefore, the mean-field approximation leads to the effective Hamiltonian for the electrons,
\begin{equation}\label{Hel}
H_{el} = \sum_{\alpha=L,R} H_{\alpha} + H_{T} + \Delta d^{\dagger}d + \lambda\expct{a+a^{\dagger}}{} d^{\dagger}d
\end{equation}
and the phonon Hamiltonian can be written as
\begin{equation}
H_{ph} = \Omega a^{\dagger} a+\lambda \expct{ d^{\dagger}d}{} (a+a^{\dagger}).
\end{equation}
In terms of the dimensionless position and momentum operator $q = \tfrac{1}{\sqrt{2}} (a+a^{\dag})$ and $p = \tfrac{i}{\sqrt{2}}(a^\dag-a)$, we can rewrite the phonon Hamiltonian as
\begin{equation}
H_{ph} = \frac{\Omega}{2} \left(q^2+p^2\right) + \sqrt{2}\lambda n q - \frac{\Omega}{2}
\end{equation}
where the electron density is denoted as $n=\expcts{d^{\dagger}d}{}$. Hence, we see that the electron-phonon coupling leads to a shift of the rest position to 
\begin{equation}
 \expct{q}{} = - \frac{\sqrt{2}\lambda n}{\Omega}\ .
\end{equation} 
In the adiabatic case, this equilibrium position follows the time evolution of $n(t)$ and thus changes on the time scales $\Delta^{-1}$ and $\Gamma^{-1}$. If the electronic dynamics is slower than the phonon dynamics, the physical picture is that of a very fast oscillator that slowly adjusts its equilibrium position. Inserting this shift into Eq.~(\ref{Hel}), we find that the back-action from the mean displacement leads to a time-dependent dot energy
\begin{equation}\label{Delta_t}
\Delta'(t) = \Delta - \frac{2\lambda^2}{\Omega}n(t).
\end{equation}
After this motivation, we shall derive the real-time dynamics of the system in this regime in a more formal manner. The diagrammatic access to the mean-field ansatz is gained by a summation of tadpole terms, which is justifiable for large $\Omega$, since the perturbation theory reveals that the tadpole term is the dominant one in this case. For this purpose, we start from Eq.~(\ref{D2tadpole}) and replace two dot GFs on the right hand side by their exact counterparts. Thus, we obtain the Dyson equation,
\begin{align}
D_{\text{mf}}(s,s') &= 
D^{(0)}(s,s') + 2\lambda^2 \int_\C ds_1\int_\C ds_2\ D^{(0)}(s,s_1)  \notag \\ &\times D_{\text{mf}}(s_1,s') F( s_1-s_2) n_{\text{mf}}(s_2)
\end{align}
This can be written in the conventional form ${\bf D} = {\bf D}^{(0)} + {\bf D}^{(0)} \Sigma {\bf D}$ by defining a self-energy in Keldysh space. For short phonon periods and coherence times compared to $\Gamma^{-1}$ and $\Delta^{-1}$, it can be approximated as
\begin{align}
 \Sigma_{\text{mf}}(s_1,s_2) 
&= 
 - \frac{2 \lambda^2}{\Omega} n_{\text{mf}}(s_1) \delta(s_1 - s_2)
\end{align} 
This time-dependent self-energy corresponds to a time-dependent dot energy level as in Eq. (\ref{Delta_t}). We can solve this mean-field Dyson equation and end up with the following self-consistency equation for the dot population
\begin{widetext}
\begin{equation} \label{n_mf_t}
n_{\text{mf}}(t) = n^{(0)}(t) + \frac{2\Gamma \lambda^2}{\pi \Omega} \theta(t) e^{-2\Gamma t} \int^\infty_{-\infty} d\omega \frac{\sum_\alpha f_\alpha (\omega-\Delta) - 2 n_0}{\Gamma^2 + \omega^2}\left\{ \frac{\lambda^2}{\Omega}\left| h(t,\omega) \right|^2 - \imag \left[g(t,\omega) h^*(t,\omega) \right] \right\}
\end{equation}
\end{widetext}
where $n^{(0)}(t)$ is given in Eq.~(\ref{n0t}), $n_0 \in \{0,1\}$ is the initial dot population and
\begin{align}
g(t,\omega) & = \left[ e^{(i\omega + \Gamma)t} -1 \right] e^{i\frac{2\lambda^2}{\Omega}N(t)} \\
h(t,\omega) & = \int_0^t ds\ g(s,\omega) n_{\text{mf}}(s)
\end{align}
and $N(t) = \int_0^t ds\ n_{\text{mf}}(s)$. Note that Eq. (\ref{n_mf_t}) incorporates the memory behavior of the mean-field ansatz since the electron density at time $t$ directly depends on all previous values after the tunnel switching at $t=0$. This retardation makes this problem solvable numerically by a discretisation of the time axis.

\begin{figure}[t]
  \centering
  \includegraphics[width = 0.48 \textwidth]{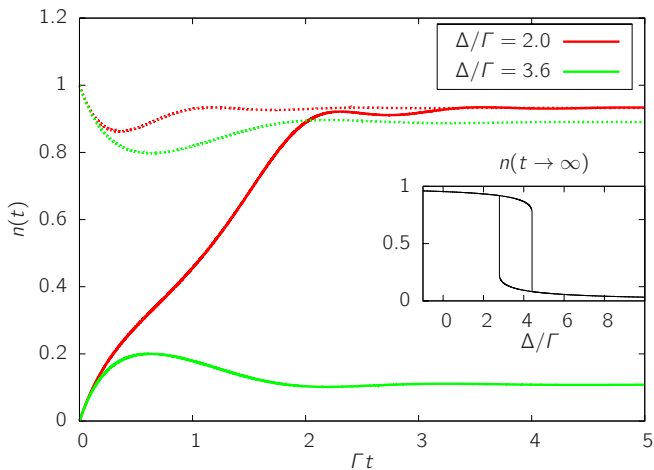}
  \caption{(Color online) Time-dependent dot population $n(t)$ for initially empty (solid lines) and occupied (dashed lines) dot level. In the bistable regime, the different initial states evolve into different stationary states. \emph{In the inset: } Bistable regions as a function of dot detuning $\Delta$. The parameters used are $\Gamma = 1, \lambda = 6$ and $\Omega = 10$.}
  \label{fig:meanfield}
\end{figure}

The resulting dot population in the mean-field approximation is shown in Fig.~\ref{fig:meanfield}. The locations of bistability as a function of $\Delta$ agree with those found in the self-consistent Hartree approximation of Ref.~[\onlinecite{damico08}]. Both stable solutions can be reached depending on the initial dot population. For $n_0 = 0$, the lower state is reached, while for $n_0 = 1$, the system evolves into the stable state with the higher dot population.

In the limit of weak electron-phonon coupling, we compared the mean-field results to the perturbative results of the previous section. It turns out that for small $\lambda$, $n_{mf}(t)$ coincides with the function resulting from the tadpole term, $\smash{n^{(2)}_1(t)}$, given in Eq.~(\ref{n2tadpole}). This is not surprising since the mean-field approach corresponds to a summation of all diagram of the tadpole type. Since for a fast phonon, the tadpole term is the  dominant one, the mean-field result reproduces the perturbative result $\smash{n^{(2)}_1(t) + n^{(2)}_2(t)}$ very well. However, no bistability exists in this perturbative regime. Conversely, perturbation theory is not applicable in the bistable regime.

Qualitatively, the function $n(t)$ as calculated by the mean-field approach retains the central characteristics known from the non-interacting case. Its time-dependence is governed by an exponential growth on a time scale $\Gamma^{-1}$ with superimposed oscillations on a time scale $\Delta^{-1}$. However, the steady-state value can be strongly influenced by the presence of the phonon mode.

\section{Conclusion}

We investigated the transient effects occurring in a molecular quantum dot coupled to a single phonon mode when the tunnelling to the leads is switched on instantly. In the case of small electron-phonon coupling, we used perturbation theory to calculate the time-dependent dot occupation $n(t)$ as well as the time-dependent currents through the two contacts $I_{L,R}(t)$.

The sudden switching leads to a number of non-adiabatic effects. We found that in the regime $\Omega \gg \Gamma$, a retardation between $I_L(t)$ and $I_R(t)$ leads to the emergence of plateau structures in the time trace of the total current, which we attribute to a feature similar to electron shuttling.

In the opposite limit of a slow phonon, $\Omega \ll \Gamma$, we find a voltage-dependent overshooting of the current compared to its steady-state value. The overshooting has a maximum at the threshold voltage for inelastic processes, $\Omega = V$.

In the case of stronger electron-phonon coupling, we used a generalisation of the mean-field approximation to investigate $n(t)$. We found that this scheme correctly reproduces the known static results and the bistability emerges naturally as the two initial dot occupations evolve into different stationary states for certain parameter constellations.

\acknowledgments
The authors would like to thank A.~Komnik and J.~Splettstoesser for interesting discussions, and in particular C.~Bruder for his constant feedback. This work was financially supported by the Swiss NSF and the NCCR Nanoscience.

\bibliography{PolaronSwitching}

\end{document}